\documentstyle[twocolumn,aps,eqsecnum,epsfig]{revtex}

\begin{document}
\draft
\preprint{CARRACK/2000-01}
\title{A coupled microwave-cavity system in the Rydberg-atom cavity
detector\\ for
dark matter axions
}
\author{M. Tada, Y. Kishimoto, M. Shibata, K. Kominato, I. 
Ogawa\cite{byline},\\
H. Funahashi$^{\rm 1}$, K. Yamamoto$^{\rm 2}$, and S. Matsuki}
\address{
Nuclear Science Division, Institute for Chemical Research, Kyoto
University, Gokasho, Uji, Kyoto 611-0011, Japan\\
${\rm 1}$ Department of Physics, Kyoto University, Kyoto 606-8503, Japan\\
${\rm 2}$ Department of Nuclear Engineering, Kyoto
University, Kyoto 606-8501, Japan
}
% \date{\today}
\maketitle
\begin{abstract}
A coupled microwave-cavity system of cylindrical TM$_{010}$ single-mode has
been developed
to search for dark matter axions around 10 $\mu {\rm eV}$(2.4 GHz)
with the Rydberg-atom cavity detector at 10 mK range temperature.
One component of the coupled cavity (conversion cavity) made of
oxygen-free high-conductivity copper is used to convert
an axion into a single photon with the Primakoff process in the strong
magnetic field,
while the other component (detection cavity) made of Nb is utilized to
detect the
converted photons with Rydberg atoms passed through it
without magnetic field.

Top of the detection cavity is attached to the bottom flange of the mixing
chamber of a
dilution refrigerator, thus the whole cavity is cooled down to 10 mK range
to reduce the
background thermal blackbody-photons in the cavity.

The cavity resonant frequency is tunable over $\sim$ 15$\%$ by moving 
dielectric rods inserted independently into each part of the cavities 
along the cylindrical axis.
In order to reduce the heat load from the higher temperature side to
the most cooled dilution refrigerator part, the tuning rod at the
conversion cavity is especially driven via the Kevlar strings
with a stepping motor outside the cryostat at room temperature.

The lowest temperature achieved up to now is 12 $\sim$ 15 mK for the
long period operation and the loaded
Q value at low temperature
is 3.5 $\sim$ 4.5 $\times$ $10^4$ for the whole range of frequency tuning.
Characteristics and the performance of the coupled-cavity system are
presented and
discussed with possible future improvements.
\end{abstract}

% \pacs{
% {\tt$\backslash$\string pacs\{\}}}

\narrowtext

\section{Introduction\protect\\ }
\label{sec:level1}

Disclosing the mystery of dark matter in the Universe is one of
the most important and challenging issues in cosmology and particle 
physics~\cite{idm98}.
A pseudo-scaler particle called axion is one of the most promising
candidates for the cold dark matter. The axion was originally
proposed to solve the so called $"$strong CP problem$"$ in the QCD
theory~\cite{Axion,Invisible}, thus being well motivated from 
the present particle physics.
Astrophysical and cosmological analyses on the contribution of
axions to the evolution of stars and also to the
matter density of the Universe can constraint the mass
of axions and the mass window still open now is from $ \sim 10^{-3}$eV
to $\sim 10^{-6}$eV~\cite{Raffelt,Kolb-Turner}.
     From the theoretical analyses of axion productions in the early
Universe, the most probable axion mass, assumed to be the dominant
component of the dark matter, is suggested to be $\sim$ 10 
$\mu$eV.
 
One of the most efficient way to search for dark matter axions is to
convert an axion into a single photon via the Primakoff process
in a high quality-factor ($Q$) microwave cavity under the
strong magnetic field and then to measure
the excess microwave power in the cavity as a function of the cavity
resonant frequency (corresponding to the mass of axions)~\cite{Sikivie}. 
Pioneering
experiments along this line were first performed by 
Rochester-BNL-FNAL group~\cite{RBF}
and then by Florida group~\cite{Hagmann1} with
a cryogenic microwave amplifier and the super-heterodyne method. The
extension of the
method has been further developed with a large scale superconducting
magnet and a cavity system by a USA group~\cite{Hagmann2}. 

\begin{figure}
\epsfig{file=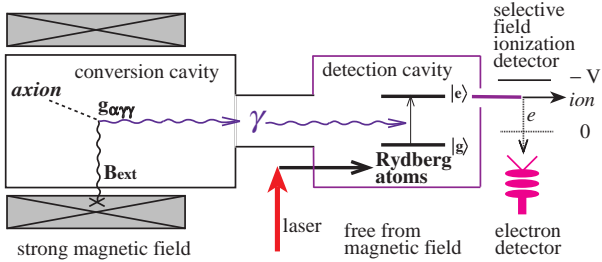, width=8cm}
\caption{Principle of the present experimental method to search
for cosmic axions with Rydberg atoms in cooled resonant cavities. The
axions are converted into photons in the conversion cavity permeated by a
strong magnetic field, and then the converted photons are absorbed by the 
Rydberg atoms in the detection cavity which is free from the magnetic field.  
Only the
excited Rydberg atoms are ionized and detected with the  selective field
ionization method. The cavities are cooled down to $\sim$ 10 mK  with
a dilution refrigerator.
}
\label{fig:principle}
\end{figure}

We have developed a Rydberg-atom single-photon 
detector~\cite{Matsuki1,Ogawa,Matsuki2} for the dark matter 
axion search.  The
experimental principle of the method is schematically shown in
Fig.~\ref{fig:principle}.   The
axion-converted photons in the resonant cavity are absorbed by Rydberg
atoms in a beam which are exclusively prepared to a lower state and the
transition frequency of which to some upper state is
approximately set equal to the cavity resonant frequency.  Passed through,
and just after exiting the cavity, the Rydberg atoms excited by absorbing
photons
are selectively ionized with the field ionization 
method~\cite{Gallagher}. Since the
Rydberg atoms are prepared only to the lower state by the multistep laser
excitation with narrow bandwidth single-mode lasers, this detection
system is almost free from the inherent noise.
Then the noise of this detection system and thus ultimate sensitivity is
mainly determined by the background from the thermal blackbody radiations
in the
cavity which can be reduced by cooling the cavity and detection
system as cold as possible with existing methods.

The essential requirements to be fulfilled for the detection system
are thus the following:
1) the system should consists of coupled microwave cavities, one
component of
which is under the strong magnetic field, while the other component
is free from the magnetic field in its inside to avoid its strong
effect on the properties of the Rydberg atoms to be used.
2) the resonant frequency of the cavity should be continuously tunable
over some range of frequency to be able to search for axions covering 
certain range of its mass.
3) the cavity and the detection system should be cooled enough to
reduce the thermal blackbody photons from the cavity wall, possibly
down to 10 mK range.

In addition to these fundamental requirements, the quality factor
($Q$) should be as high as possible and also the conversion efficiency
of axions into photons in the cavity should be as high as possible in
order to make the efficiency of axion detection high enough.

%
%
%   eps figure   *******************************
%   eps figure   *******************************
%

A pilot experimental apparatus of this line called CARRACK1 has been
developed
and is being used to search for dark matter axions at around 2.4 GHz (axion
mass of 10 $\mu$eV). To realize the above mentioned requirements
actually,  two single-mode cylindrical TM$_{010}$ cavities are coupled
through a ring-shape hole between them.  One component of the coupled
cavity is made of copper which is permeated by the strong magnetic
field produced with a superconducting magnet and the other component
is made of niobium to expel out the external magnetic field with the
Meissner effect. Specifically the cavity system is attached to the
bottom plate of the mixing chamber of a dilution refrigerator (DF) and thus
the whole cavity and the adjacent detector components are cooled down to
10 mK range in order to reduce the thermal blackbody photons in the cavity.
The cavity resonant frequency is tunable over
$\sim$ 15$\%$ by moving dielectric rods inserted independently in both cavity
components along the cylindrical symmetry axis.  In due course of the
development, much attention has been especially paid to cool the system down to
10 mK range and also to get high $Q$ cavities.

Although the present cavity system
was constructed exclusively to be dedicated to search for dark matter
axions, we believe that the underlying development of such system will 
also be
useful to other kind of applications in general.
In this note, the design principle and the actual apparatus of the
present cavity system are described in sections 2 and 3. Then the
characteristics and
the performance of the system is presented and discussed in sections 4
and 5 with possible improvements in future.  Section 6 is devoted
to summarize the results of the present investigation.

\section{Design principle\protect\\ }
\label{sec:design-principle}
\subsection{General}
\label{sec:design-general}

Assuming the dark matter of our own galaxy (dark halo) consisting of
axions, the number density of axions is given by
\begin{equation}
{\bar n}_a = 3.0 \times 10^{13}
\left( \frac{\rho_a}{0.3 {\rm GeV} {\rm cm}^{-3}} \right)
\left( \frac{10^{-5} {\rm eV}}{m_a} \right),
\label{eqn:na_dens}
\end{equation}
%%%%%
where $m_a$ is the mass of axion and the energy density of the cosmic axions
$ \rho_a $  is taken to be equal to that of the galactic dark halo
$ \rho_{\rm halo} \simeq 0.3 {\rm GeV} {\rm cm}^{-3} $.

In the following we will firstly estimate the signal-to-background ratio ($s/n$) 
of axion detection in a very crude approximation without taking into 
account the quantum nature of the Rydberg-atom cavity detector.     
The pseudoscaler axion of spin-parity 0$^-$ is converted to a photon in the
strong
magnetic field with the Primakoff process. The conversion rate in a resonant
cavity is approximately given by~\cite{Sikivie}
\begin{equation}
R = \left( \frac{\epsilon_0}{\hbar^2} \right) g_{a \gamma \gamma}^{2}
\omega_{c}^{-1} Q_{c} B_{0}^{2} G^{2},
\label{eqn:conv_rate}
\end{equation}
where $g_{a\gamma\gamma}$, $\omega_{c}$, $Q_c$, $B_0$, and $G$ are the
axion-photon coupling constant, the cavity resonant angular frequency,
the cavity quality factor, maximum magnetic flux density and the
geometric form factor of the cavity of order unity as described later
in detail, respectively.

Taking into account the volume of the cavity, to be of order 10$^3$ cm$^3$
for axions with mass of 10 $\mu$eV, the number of converted photons produced by
the Primakoff process is estimated to be of order 0.1 to 1 from the above
equations
with conventionally available superconducting magnet.  While this number
seems to be
significant, yet the background thermal photon number is much larger
than this number even at 4 K temperature: In fact the mean number of
thermal blackbody radiations $ {\bar n}_c $ present in a resonant 
single-mode cavity is given by 

%%%%%
\begin{equation}
{\bar n}_c
= \left( {\rm e}^{\hbar \omega_c / k_{\rm B} T_c} - 1 \right)^{-1} .
\label{eqn:nbarc}
\end{equation}
%%%%%
 
     From this mean number, the number of background photons detected by a
detector with effective $Q$ value of $Q_{det}$ is given by

\begin{equation}
N_d = {\bar n}_c \gamma_c \left( \frac{Q_{det}}{Q_{c} + Q_{det}} \right),
\label{eqn:N_d}
\end{equation}

where $\gamma_c$ is the dumping factor of the cavity given by

%%%%%
\begin{eqnarray}
\gamma_c \equiv \frac {\omega_c}{Q_c} & = & 6.6 \times 10^5 \left( \frac{f_c}{2.4 {\rm
GHz}} \right)
\left( \frac{3 \times 10^4}{Q_c} \right).
\label{eqn:gamma}
\end{eqnarray}
%%%%%

The $s/n$ ratio in a second is thus given
approximately by
\begin{eqnarray}
s/n & \simeq & O(1) \left( \frac{10 \mu{\rm eV}}{m_a} \right)^{2} \left(
\frac{Q_c}{3 \times 10^4} \right)^{2} \left( \frac{B_{0}}{7{\rm T}} \right)^{2}
\nonumber \\
& \times & \left( \frac{10^{5 \cdot \left( \frac{m_a}{10 \mu {\rm eV}} \right) \left( 
\frac{10 {\rm mK}}{T} \right)}}{10^5} \right),
\label{eqn:s/n}
\end{eqnarray}
where it is assumed that $k_{\rm B}T \ll {\hbar \omega_c} $. 
     From this equation, it is clear that the cavity has to be cooled down to
10 mK range if the $s/n$ ratio is required to be of order better than one.

\subsection{Quantum treatment of axion - photon - Rydberg-atom
interactions in resonant cavities}
\label{sec:quantum_theory}

Although the above analyses are useful for the very crude estimation of
the detection efficiency, more rigorous analyses have to be based on
the quantum theory of axion-photon-atom interactions as clarified
previously~\cite{Ogawa,Yamamoto,Kitagawa} in detail. 
Theoretical analyses are briefly discussed in the following.

The axion-photon interaction under a strong static magnetic field
with flux density $ {\mbox{\boldmath $ B $}}_0 $
is described by the Lagrangian density
%%%%%
\begin{equation}
{\cal L}_a = \hbar^{1/2} \epsilon_0 g_{a \gamma \gamma}
\phi {\mbox{\boldmath $ E $}} \cdot {\mbox{\boldmath $ B $}}_0 ,
\label{eqn:La}
\end{equation}
%%%%%
where $ \hbar^{1/2} $ and $ \epsilon_0 $ (dielectric constant)
are explicitly factored out
so that the Lagrangian density has the right dimension
$ {\cal L}_a \sim \hbar {\rm s}^{-1} {\rm m}^{-3}
\sim {\rm eV} {\rm m}^{-3} $
with the axion-photon-photon coupling constant
$ g_{a \gamma \gamma} \sim {\rm eV}^{-1} $.
The axion-photon-photon coupling constant is 
calculated~\cite{Axion,Invisible,KS} 
as
%%%%%
\begin{equation}
g_{a \gamma \gamma} = c_{a \gamma \gamma}
\frac{\alpha}{2 \pi^2} \frac{m_a}{f_\pi m_\pi} \frac{(1 + Z)}{{\sqrt Z}} ,
\label{eqn:gagg}
\end{equation}
%%%%%
where $ Z = m_u / m_d $, and
%%%%%
\begin{equation}
c_{a \gamma \gamma} = \frac{D}{C} - \frac{2(4 + Z)}{3(1 + Z)}
\end{equation}
%%%%%
with
%%%%%
\begin{equation}
D = {\rm Tr} Q_{\rm PQ} Q_{\rm em}^2 , \
C \delta_{ab} = {\rm Tr} Q_{\rm PQ} \lambda_a \lambda_b .
\end{equation}
%%%%%
The parameter $ c_{a \gamma \gamma} $ represents
the variation of the axion-photon-photon coupling
depending on the respective Peccei-Quinn models such as
the so-called KSVZ
and DFSZ~\cite{Invisible}.

The electric field operator in the cavity $ {\cal V} $ is given by
%%%%%
\begin{equation}
{\mbox{\boldmath $ E $}} ({\bf x},t)
= ( \hbar \omega_c / 2 \epsilon_0 )^{1/2}
[ {\mbox{\boldmath $ \alpha $}} ({\bf x}) c(t)
+ {\mbox{\boldmath $ \alpha $}}^* ({\bf x}) c^{\dagger}(t) ]
\label{eqn:E}
\end{equation}
%%%%%
for the radiation mode with a resonant frequency $ \omega_c $,
where $ \epsilon_0 $ is the dielectric constant.
The creation and annihilation operators satisfy
the usual commutation relation
%%%%%
\begin{equation}
[ c , c^\dagger ] = 1 .
\label{eqn:ccdgg}
\end{equation}
%%%%%
The mode vector field $ {\mbox{\boldmath $ \alpha $}}({\bf x}) $
is normalized by the condition
%%%%%
\begin{equation}
\int_{\cal V} | {\mbox{\boldmath $ \alpha $}} ({\bf x}) |^2 d^3 x = 1 .
\label{eqn:a-nrm}
\end{equation}
%%%%%
The whole cavity $ {\cal V} $
may be viewed as a combination of two subcavities,
the conversion cavity $ {\cal V}_1 $ with volume $ V_1 $
and the detection cavity $ {\cal V}_2 $ with volume $ V_2 $,
which are coupled together:
%%%%%
\begin{equation}
{\cal V} = {\cal V}_1 \oplus {\cal V}_2 .
\end{equation}
%%%%%
The axion-photon conversion takes place in $ {\cal V}_1 $
under the strong magnetic field,
while the Rydberg atoms are excited by absorbing the photons
in $ {\cal V}_2 $.  It is then suitable to divide the mode vector as
%%%%%
\begin{equation}
{\mbox{\boldmath $ \alpha $}} ({\bf x})
= {\mbox{\boldmath $ \alpha $}}_1 ({\bf x})
+ {\mbox{\boldmath $ \alpha $}}_2 ({\bf x}) ,
\label{eqn:modev}
\end{equation}
%%%%%
where $ {\mbox{\boldmath $ \alpha $}}_1 ({\bf x}) = {\bf 0} $
for $ {\bf x} \in {\cal V}_2 $
and $ {\mbox{\boldmath $ \alpha $}}_2 ({\bf x}) = {\bf 0} $
for $ {\bf x} \in {\cal V}_1 $, respectively.
The normalization condition of $ {\mbox{\boldmath $ \alpha $}} ({\bf x}) $
is rewritten as
%%%%%
\begin{equation}
\int_{{\cal V}_1} | {\mbox{\boldmath $ \alpha $}}_1 ({\bf x}) |^2 d^3 x
+ \int_{{\cal V}_2} | {\mbox{\boldmath $ \alpha $}}_2 ({\bf x}) |^2 d^3 x
= 1 .
\label{eqn:moden}
\end{equation}
%%%%%

The actual cavity is designed so that neglecting the small joint region
the subcavities $ {\cal V}_1 $ and $ {\cal V}_2 $ admit the mode vectors
$ {\mbox{\boldmath $ \alpha $}}_1^0 ({\bf x}) $
and $ {\mbox{\boldmath $ \alpha $}}_2^0 ({\bf x}) $
(up to the normalization and complex phase), respectively,
whose frequencies are tuned to be almost equal to some common value
$ \omega_c^0 $.
In this situation, as confirmed by numerical calculations
and experimental observations,
two nearby eigenmodes with the frequencies
$ \omega_c ,\ \omega_c^\prime \simeq \omega_c^0 $ are obtained
for the whole cavity $ {\cal V} $.  Then, the mode vector
$ {\mbox{\boldmath $ \alpha $}} ({\bf x}) $ is constructed
approximately of
$ {\mbox{\boldmath $ \alpha $}}_1 ({\bf x})
\simeq {\mbox{\boldmath $ \alpha $}}_1^0 ({\bf x}) $
and $ {\mbox{\boldmath $ \alpha $}}_2 ({\bf x})
\simeq {\mbox{\boldmath $ \alpha $}}_2^0 ({\bf x}) $
with significant magnitudes in both $ {\cal V}_1 $ and $ {\cal V}_2 $.
The conversion of the cosmic axions takes place predominantly
to the radiation mode which is resonant with the axions
satisfying the condition
$ | \omega_c - m_a / \hbar | \lesssim \gamma_a $
(axion width) $ \sim $ small fraction of $ \gamma_c $ (cavity damping rate).
The cavity can be designed so as to give a sufficient separation
of $ | \omega_c - \omega_c^\prime | > {\rm several} \ \gamma_c $
for the nearby modes with strong coupling
between $ {\cal V}_1 $ and $ {\cal V}_2 $.
Therefore, in the search for the signal from the cosmic axions,
the one resonant mode can be extracted solely
for the electric field in a good approximation, as given
in Eq. (\ref{eqn:E}), whose frequency $ \omega_c $ is supposed to be
close enough to the axion frequency $ \omega_a = m_a / \hbar $.

The original Lagrangian density for the axion-photon-photon coupling
in Eq. (\ref{eqn:La}) provides the effective interaction Hamiltonian
between the coherent axion mode $ a $
and the resonant radiation mode $ c $,
%%%%%
\begin{equation}
H_{ac} = \hbar \kappa ( a^\dagger c + a c^\dagger ) .
\label{eqn:H-ac}
\end{equation}
%%%%%

     The coherent axion mode can be identified as
%%%%%
\begin{equation}
a(t) = \Sigma_a^{-1/2}
\int_{{\cal R}_a} \frac{d^3 k}{(2 \pi )^3 2 \omega_k} \ a_{\bf k} (t) .
\label{eqn:a(t)}
\end{equation}
%%%%%
where $R_a$ represents the coherent region of axions, and the normalization
factor is given by
%%%%%
\begin{equation}
\Sigma_a = \int_{{\cal R}_a} \frac{d^3 k}{(2 \pi )^3 2 \omega_k}
\equiv \frac{1}{2m_a} \left( \frac{\beta_a m_a}{2 \pi \hbar} \right)^3 ,
\label{eqn:sigma-a}
\end{equation}
%%%%%
so that the coherent mode operator satisfies
the canonical commutation relation,
%%%%%
\begin{equation}
[a, a^{\dagger}] = 1 .
\label{eqn:a-comm}
\end{equation}
%%%%%

%
%
%   eps figure   *******************************
\begin{figure}
\epsfig{file=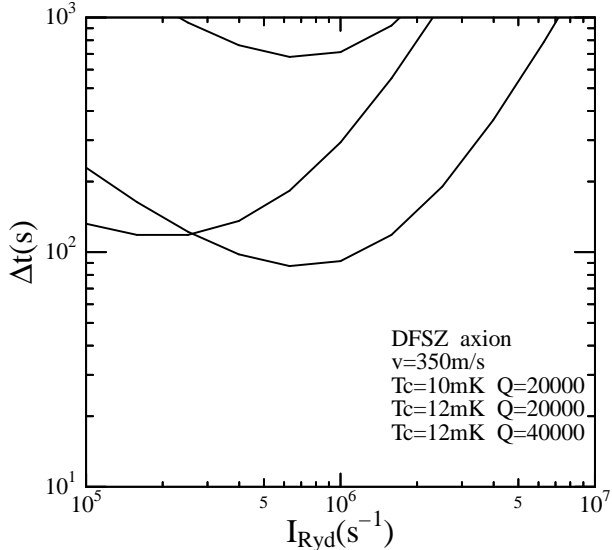, width=8cm}
\caption{ Necessary time to observe the axion-converted photon signals 
in 3$\sigma$ level as a function of the average number of Rydberg atoms 
in resonant cavity under the existence of the background thermal 
blackbody photons.  The axion-converted photon signals are estimated 
from the DFSZ axion model.
}
\label{fig:delta-t}
\end{figure}
%   eps figure   *******************************
%

The axion-photon conversion in the cavity $ {\cal V}_1 $
is well described with this interaction Hamiltonian.
The coupling constant $ \kappa $ is determined
for $ \omega_c \simeq m_a / \hbar $ ~\cite{Axion,Invisible,KS}
%%%%%
\begin{eqnarray}
\kappa & = & \hbar^{1/2} g_{a \gamma \gamma} \epsilon_0^{1/2}
B_{\rm eff} \left[ \left( \frac{\beta_a m_a}{2 \pi \hbar} \right)^3
\frac{V_1}{2} \right]^{1/2}
\nonumber \\
& = & 4 \times 10^{-26}{\rm eV} \hbar^{-1}
\left( \frac{g_{a \gamma \gamma}}{1.4 \times 10^{-15}{\rm GeV}^{-1}} \right)
\left( \frac{B_{\rm eff}}{4{\rm T}} \right)
\nonumber \\
& \times & \left( \frac{\beta_a m_a }{10^{-3}
\times 10^{-5}{\rm eV}} \right)^{3/2}
\left( \frac{V_1}{5000{\rm cm}^3} \right)^{1/2} ,
\label{eqn:kappa}
\end{eqnarray}
%%%%%
where
%%%%%
\begin{equation}
B_{\rm eff} = \zeta_1 G B_0 ,
\end{equation}
%%%%%
and $ B_0 $ is the maximal density of the external magnetic flux.
The axion-photon-photon coupling constant $ g_{a \gamma \gamma} $
is taken here to be the value expected from the DFSZ axion 
model~\cite{Invisible} at $ m_a = 10^{-5} {\rm eV} $.

The form factor for the magnetic field is given by
%%%%%
\begin{equation}
G = \zeta_1^{-1} V_1^{-1/2} \left| \int_{{\cal V}_1} d^3 x
{\mbox{\boldmath $ \alpha $}}_1 ({\bf x}) \cdot
[ {\mbox{\boldmath $ B $}}_0 ({\bf x})/B_0 ] \right|
\label{eqn:G}
\end{equation}
%%%%%
with
%%%%%
\begin{equation}
\zeta_1 = \left[ \int_{{\cal V}_1} d^3 x
| {\mbox{\boldmath $ \alpha $}}_1 ({\bf x}) |^2 \right]^{1/2} .
\label{eqn:zeta1}
\end{equation}
%%%%%
This additional factor $ \zeta_1 $ ($ < 1 $
as seen from Eq.(\ref{eqn:moden}))
represents the effective reduction of the axion-photon conversion
which is due to the fact that the magnetic field
is applied only in the conversion cavity $ {\cal V}_1 $.
We may obtain, for example, the effective magnetic field strength
$ B_{\rm eff} \simeq 4 {\rm T} $,
as taken in Eq.(\ref{eqn:kappa}),
by using typically a magnet of $ B_0 \simeq 7{\rm T} $
and the cavity system with $ G = {\sqrt{0.7}} $
of $ {\rm TM}_{010} $ mode and $ \zeta_1 \simeq 0.7 $
for the conversion cavity. The actual value of $B_{\rm eff}$ was evaluated
with the calculated electric field distributions in the cavity as
described in detail later.

Now the time evolution of the axion - photon - Rydberg-atom system is governed
by the following equations of motion in the Heisenberg picture,
when all of the Rydberg atoms are prepared initially in the lower 
state~\cite{Ogawa,Yamamoto,Kitagawa}:
%%%%%%%%%%%%%%%%%%%%
\begin{equation}
\frac{dz_i}{dt} = K_{ij} z_j + F_i ~,
\label{eqn:dz/dt}
\end{equation}
%%%%%%%%%%%%%%%%%%%%
where $ z_i = ( b , c , a ) $, $ F_i = ( 0 , F_c , F_a  ) $, and
%%%%%%%%%%%%%%%%%%%%
\begin{equation}
K = \left( \begin{array}{ccc}
- i \omega_b & i \Omega_N & 0 \\
i \Omega_N & - i \omega_c - \frac{1}{2} \gamma_{c} & i \kappa \\
0 & i \kappa & - i \omega_a - \frac{1}{2} \gamma_a  \\
\end{array} \right)
\label{eqn:K}
\end{equation}
%%%%%%%%%%%%%%%%%%%%
with  $ \gamma_a \simeq \beta_a^2 m_a $, the energy dispersion of 
the cosmic axions. The operators $a$,$b$, and $c$ refer to axion, atom 
and cavity-mode photon, respectively.    
The external forces $ F_c $ and $ F_a $ are introduced
for the Liouvillian relaxations of the photons and axions, respectively
\cite{Haroche,Louisell}.

By solving the equations, the detection efficiency can be numerically
calculated as a function of various experimental parameters such as
the transit time of the atomic beam in the detection cavity, cavity
temperature and the number of Rydberg atoms passed through the cavity.
  From these results the optimum condition of the experimental
parameters like the transit time of the Rydberg atoms, or the velocity
of the atoms can be determined. The cavity system was thus designed to meet
these requirement. 

Along the line of these analyses, further refined treatment of the system
has been recently performed by taking into account the spatial distribution of
the Rydberg atoms in the cavity and also the actual electric field
distribution along the path of the atoms in the cavity. The
resulting modifications were also taken into account for the design
of the cavity system, although the modifications are not 
profound~\cite{Yamamoto,Kitagawa}. 

 Specifically the most important ingredient among 
these parameters is the average number of Rydberg atoms present in 
the cavity.  Due to the collective interactions of the atoms with the 
cavity mode, the overall coupling strength is proportional to the square of the 
number of atoms in the cavity so that the most optimum coupling 
strength can be tuned by adjusting the injecting number of Rydberg 
atoms, thus with the oven temperature of thermal atomic beam and/or 
with the power of the second laser.     

Typical dependence of the necessary time to search for axions over 10 
\% region with 3$\sigma$ level on the number of the Rydberg atoms are 
shown in Fig.~\ref{fig:delta-t}. From this dependence it is found 
that the optimum number of Rydberg atoms should be around $5 
\times 10^{5} {\rm sec}^{-1}$ for the 10 $\mu$eV axions with the mean atomic 
velocity of 350 m/sec.        

%
%
%   eps figure   *******************************
\begin{figure}
\epsfig{file=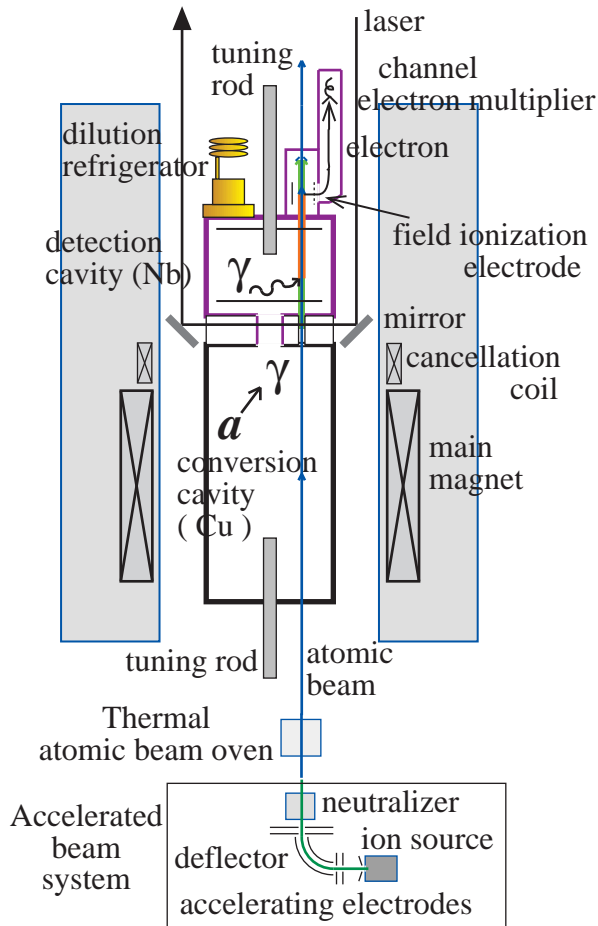, width=8cm}
\caption{Schematic diagram of the present experimental system to search
for cosmic axions with Rydberg atoms in cooled resonant cavities.  
The cavities and the related detector system are all cooled down to $\sim$ 
10 mK with a dilution refrigerator to reduce the thermal blackbody 
photons.  Accelerated or thermally produced atomic beam of Rb in its 
ground state is utilized for the Rydberg atoms.
}
\label{fig:setup}
\end{figure}
%   eps figure   *******************************
%

\subsection{Rydberg-atom cavity detector}
\label{sec:design-detector}

Following the above theoretical analyses of the optimum setup of the 
experimental system, the whole system of the present axion search 
apparatus was designed as schematically shown in Fig.~\ref{fig:setup}.  
Thermal or accelerated 
atomic beam, produced from a thermal Rb oven placed beneath the 
cryostat, are injected 
into the low 
temperature cavity and stopped at 1K-pot plate of the dilution 
refrigerator system after passing through the detection cavity. Just in front 
of the detection cavity the ground state atoms of $^{85}$Rb are excited to       
a Rydberg state with 2- or 3-step laser excitation. Laser beams 
are introduced from the top into the cryostat and after interacting with
atoms are then extracted outside through a glass window. 

The Rydberg atoms excited to an upper state are ionized at the field 
ionization electrode and the electrons thus produced are guided 
through a series of focussing ring-electrodes to, and detected  
with, a channeltron electron multiplier placed at the 1K-pot plate. 
The whole cavity and 
the field ionization detection system are cooled down to 10 mK range 
with the DF. For this purpose the top of the detection 
cavity is attached to the bottom plate of the mixing chamber.

\section{Apparatus\protect\\ }
\begin{figure}
\epsfig{file=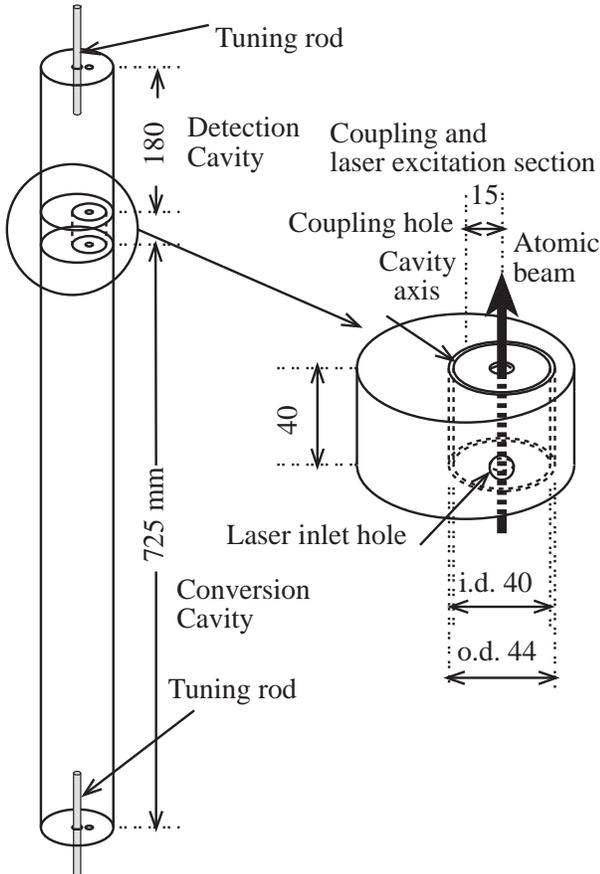, width=8cm}
\caption{Coupled cavity and coupling structure in the present cavity 
system. 
}
\label{fig:cavity}
\end{figure}

The whole coupled-cavity system is shown in Fig.~\ref{fig:cavity}.

The cavity system is located in the inner vacuum
chamber surrounded by a liquid He bath, a liquid nitrogen bath and
a outer vacuum chamber.  

Two TM$_{010}$ single-mode cavities of the
same inner diameter are connected to each other through a ring-shape
hole. The resonant frequency of the cylindrical TM$_{010}$ cavity is 
determined by its diameter $D$ as 

\begin{equation}
f_{c} \equiv \frac {\omega_{c}} {2\pi} = 2.55 {\rm GHz} \left( \frac{90 {\rm 
mm}}{D} \right).   
\label{eqn:freq}
\end{equation}

Two cavity components are made of oxygen-free high-conductivity
copper OFHC (conversion
cavity) and niobium (detection cavity), respectively. The detection cavity 
is made of superconducting niobium to
expel out the external magnetic field with the Meissner effect.

The tuning of the cavity resonant frequency
is accomplished by inserting and moving dielectric rods in both
the conversion and the detection cavities along the cylindrical
symmetry axis. This choice of the tuning method was adopted from its simple
mechanism
of driving from the outside of the cryostat with stepping motors,
although the form factor of the cavity is not the best. The rod for
the detection cavity is driven from the top of the cryostat, while
that for the conversion cavity is driven from the bottom of the
cryostat.

%
%
%   eps figure   *******************************
%   eps figure   *******************************
%

\subsection{coupled cavity}
\label{sec:coupled-cavity}

The two cavity components are coupled through a ring-shape hole as
shown in Fig.~\ref{fig:cavity}. The strength of the coupling can be estimated
by the frequency separation of the two eigen modes. This separation
is related to how fast the stored power in one component of the cavity is
transferred to another
component.  This frequency separation was estimated with a two-dimensional 
computer code SUPERFISH for alternating electromagnetic field
calculations. The field
distribution in the cavity is affected by the inserted
rod position, due to the large dielectric constant of the rod material.
Although the electric field direction is not always perfectly
parallel to the cylindrical axis, the whole field distribution is
consistent with the TM$_{010}$ mode for the full range of the
frequency as a function of the rod position.

One of the eigen mode (parallel mode) has an electric
field directions in parallel for both components, while in the
other eigen mode (anti-parallel mode), the field direction of the
conversion cavity is
anti-parallel to that of the detection cavity.  The electric field in the
anti-parallel mode is stronger in the conversion cavity than in the
detection cavity,
while opposite situation is realized in the parallel mode.

The anti-parallel eigen mode, in which the electric field in the conversion
cavity is
stronger, was chosen for the actual cavity system in the present case.

%
%
%   eps figure   *******************************
\begin{figure}
\epsfig{file=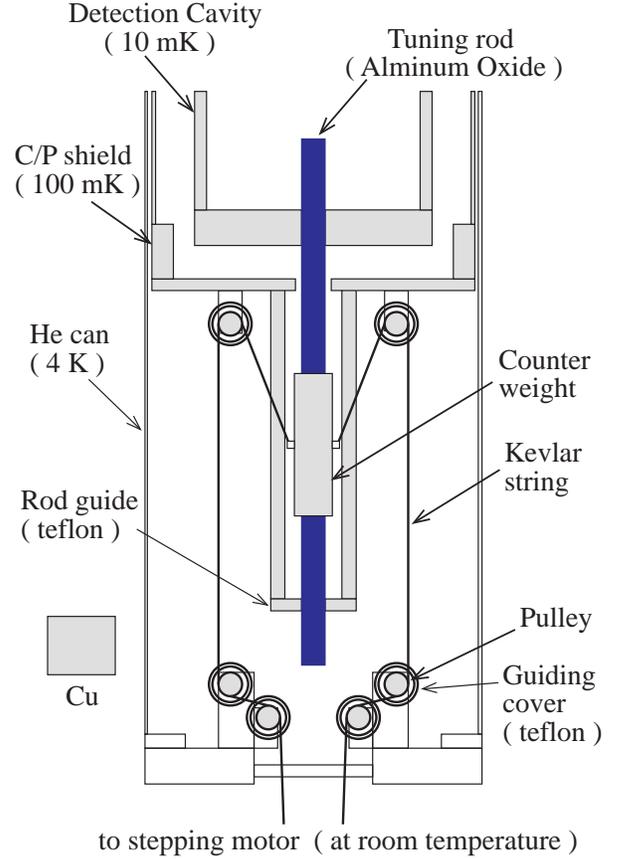, width=8cm}
\caption{Schematic drawing of the driving mechanism of the rod in the 
conversion cavity with Kevlar strings for frequency tuning.  The 
strings are moved with a stepping motor outside the cryostat at room 
temperature.
}
\label{fig:string}
\end{figure}
%   eps figure   *******************************
%

\subsection{Frequency tuning\protect\\}
\label{sec:apparatus2}
The frequency tuning is accomplished by inserting the dielectric rods
to both the detection and conversion cavities along the cylindrical
axis. Depending on the size of the inner diameter of the cavity, the
diameter of the tuning rod is determined for the resonant frequency
to cover the expected frequency regions. Aluminum-oxide rod of 7 mm
diameter is used for both cavities. 

Since the cavity is cooled down to 10 mK range, heat leak from the
higher temperature side is extremely important. For the detection
cavity, which is located at the upper part of the cavity system and
attached to the bottom plate of the mixing chamber, the rod is
inserted to the cavity through a stainless-steel pipe and a TI-polymer 
pipe~\cite{Toray} 
from the top of the cryostat.  The rod was anchored to the
liquid nitrogen shield plate, liquid He can, and then to the cold plate
shield (c/p shield plate) at $\sim$100 mK temperature with a strand 
of copper meshes.

Only the c/p shield at 100 mK is available for the rod of the
conversion cavity to be anchored so that special care has to be taken 
to avoid the
heat leak from the room temperature side coming into the cavity at 10 mK range.
The lower-side rod for the conversion cavity is thus moved via Kevlar
strings~\cite{Toray} with a
stepping motor outside the cryostat at room temperature. The
driving mechanism is shown schematically in Fig. ~\ref{fig:string}. The Kevlar
string has quite low heat
conductivity and still has enough strength to drive the rod of weight
$\sim$100 g. It has also quite low extensibility.  The heat leak from the
room temperature parts is estimated to be less than 50 nW. 

\subsection{Magnetic field shielding}
Although we need strong magnetic field in the conversion
cavity, it is easier to handle the Rydberg-atom single-photon detector
in a circumstance without magnetic field to avoid the magnetic field 
in the detection cavity, because
otherwise we have a complicated energy levels of the Rydberg states
due to the effect of the Zeeman splitting.
To expel out the strong magnetic field from the detection cavity,
following method was adopted in the region of the detection
cavity:first, the external magnetic field at the detection cavity was reduced
to less than 0.09 T with a superconducting coil (cancellation coil) with 
counter flow of the
current against the main magnet coil. Second, the detection cavity and the coupling
part are made of Nb metal, which become a superconducting state at the
cooled stage, thus the inner field being expelled out due to the
Meissner effect.  

It is noted here that in order to well realize the field free region
in the detection cavity, we have to carefully take into account the effect
of the
demagnetization field induced by the superconducting Nb metals.  With
a three-dimensional eddy current program EDDY~\cite{photon}, the magnetic flux
density resulting from the insertion of the superconducting Nb metals was
calculated and the proper shape of the Nb cavity were thus designed
to expel the flux density out of the detection cavity. The
detailed treatment of the magnetic shield will be reported elsewhere.

\subsection{Selective-field-ionization detector}
Selective field ionization detector consists of three parts; field 
ionization electrodes, transport electrodes for the ionized electrons 
and the electron multiplier for detecting the electrons. The field 
ionization electrodes are located in the Nb box set at the bottom 
plate of the mixing chamber and the Rydberg atoms entered into the 
electrodes after passing through the detection cavity. The ionized 
electrons are then transported into the electron multiplier located 
at the 1K-pot plate through the number of ring electrodes for 
focussing.  These transport electrodes are distributed from the 
vicinity of the ionization electrodes to the 1K-pot region through the 
cold/plate region to effectively transport the electrons.  

These configurations are adopted to keep the cavity and the field 
ionization parts as cool as possible, since the used channel electron 
multiplier 
produces heat power as large as 10 mW. Moreover the channeltron 
multiplier can not be used at such low temperature as 10 mK range, 
because the amplification gain is strongly reduced a t such low 
temperature.  Therefore the channeltron multiplier has to be always kept at 
temperature higher than 20 K. In order to fulfill this requirement, 
the multiplier was heated up with a heating coil surrounding it and 
separated thermally from the 1K-pot plate by supporting it with a low 
thermal conductivity dielectric material called TI polymer. The introduced 
power of the heating 
coil is about 30 mW and the current flowing between the top and the 
collector electrodes of the channeltron multiplier is about 5 $\mu$A. 
Detailed description of the SFI detector will be published elsewhere.

%
%
%   eps figure   *******************************
\begin{figure}
\epsfig{file=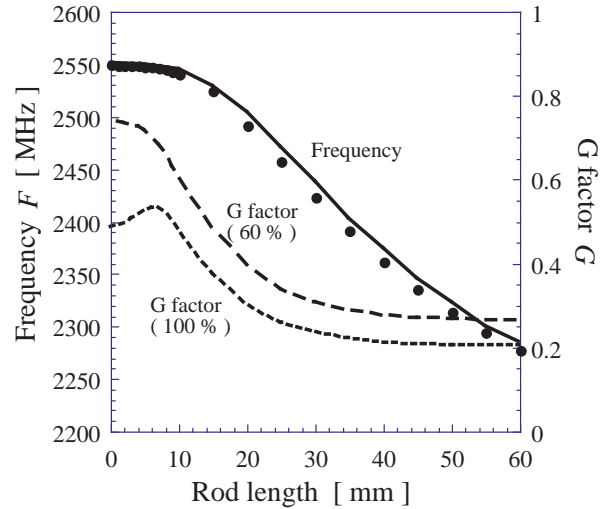, width=8cm}
\caption{Frequency and form-factor variation with the position of the 
tuning rod in 
the conversion cavity.  Solid line is the calculated frequency 
variation and the solid circles show the observed results.  Also 
shown is the $G$ factor with (labeled 60\% in the Figure where 40\% 
of the cavity at the lower part is cut off for the estimation) and 
without (denoted 100\%) taking into account 
the lower part of the conversion cavity. See text in detail for the 
treatment of the lower part of the cavity.
}
\label{fig:fvsrod}
\end{figure}
%   eps figure   *******************************
%

\subsection{Associated equipment}
In addition to the coupling hole between the two sub-cavities and the 
holes for the frequency tuning rod, each cavity has
also a hole to introduce the Rydberg atomic beam into the
cavity and to make the atoms pass through for detecting
the excited atoms with the field ionization method. These holes have
some effect on the degradation of the loaded Q value of the cavity.

Several associated equipments were also installed
to the cavity system. One is the
electrodes for inducing the Stark shift of the Rydberg states inserted
along the beam path in the cavity. The electrodes are used to tune
finely the Rydberg-atom transition frequency to match with the cavity
resonant frequency and thus to be able to search for axions over some range of
axion mass. The detailed structure of the Stark electrodes will be 
reported elsewhere.

Since the total length of the whole cavity system is 80 cm, some
acoustic vibrational motion may possibly be induced at the lower
part of the cavity.  In order to avoid this vibration, the lower part
of the cavity were connected to the c/p cylindrical shield plate with several
small graphite rods. The induced heat leak from the c/p shield to the
cavity is negligible due to the low thermal heat conductivity of the
graphite used.

\section{Performance\protect\\ }
\subsection{frequency tuning and the Q value}
 
The cavity resonant frequency was measured with a tracking analysis
system in the ADVANTEST R3261A spectrum analyzer. The coupling
antenna in both the cavity components are straight thin wire of about 1 mm
length. Semi-rigid UT141 cables made of Be/Cu were used to connect the antenna 
to the main amplifier outside of the cryostat.   
In cooled stage, a cryogenic amplifier and a circulator
inserted between the cavity and the main amplifier outside the cryostat were
used to increase the sensitivity for the calibration. However once the 
calibration measurement has been finished, these auxiliary equipments 
were removed, since they induce some noise into the cavity system and 
thus degrade the performance of the present Rydberg-atom cavity detector.

The tuning of the cavity resonant frequency was controlled by  varying the
position of the rod in the cavity with a stepping motor through a
computer.  A data acquisition and control program LabVIEW was used for 
the whole control of the tuning system.  It is noted that the tuning  
of the CC rod position is mostly effective for the cavity frequency 
tuning and is not so sensitive to the DC rod position.  It is 
therefore not necessary for the tuning to finely adjust the positions 
of the two rods relatively to each other.       

The measured frequency range covered by the tuning
rod is shown in Fig.~\ref{fig:fvsrod} together with the calculated result
for comparison.
Although the extensibility of the strings used is quite small, the
effect of backlash of the stepping motors has some effect of the
reproducibility of the resonant frequencies when back and forth
movement has been made. However these backlash effect can be avoided
by always moving the rod to the same direction during the course of the
experiment.

The loaded Q value observed is $3.5 \sim 4.5 \times 10^4$ for the whole
range of the resonant frequencies tuned.  The Q value achieved is 
quite sensitive to the goodness of the contact between the side cylinder and the 
upper and the lower plates.  In between them we inserted and tightened   
thin indium wires and/or thin copper rings which have especially sharp 
edges on both side of the connecting surfaces.  Both methods worked 
well, although the use of copper rings is more flexible in practice.    

The stability of the resonant frequency for a long period is one of the
important factors for the present purpose of the experiment. We
checked the stability of the frequency by fixing the rod position.
It was found that the resonant frequency is stable within 3 kHz for more
than 3 hours, enough stability for the present purpose.

%
%
%   eps figure   *******************************
\begin{figure}
\epsfig{file=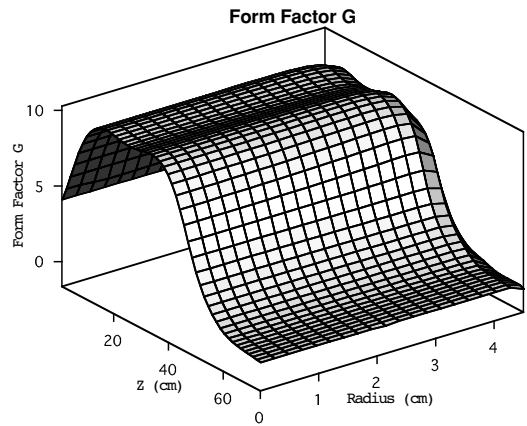, width=8cm}
\caption{Distribution of form factor $G$ in the present  
cylindrical TM$_{010}$ cavity. 
}
\label{fig:G3dim}
\end{figure}
%   eps figure   *******************************
%

\subsection{Conversion form-factor\protect\\}
\label{sec:emf-cff}
The form factor for the axion conversion is proportional to 
${\boldmath E} \dot {\boldmath B}$ as described in the previous 
section. In Fig.
~\ref{fig:fvsrod} shown is the form factor $G$ evaluated from the 
electric field distribution in the cavity with the
code SUPERFISH. The magnetic field in the cavity is
calculated from the superconducting coil dimensions with the program
POISSON.

The distribution of the $G$ factor in the cavity is shown in 
Fig.~\ref{fig:G3dim} in three dimensional representation.  The 
distribution along the symmetry z axis is mainly due to that of  
the magnetic field. 

It is noted that from the view 
point of maximum G-factor attainable, there should be some optimum 
value for the length of the conversion cavity.  However due to the 
requirement of the effective coupling to the detection cavity, the
conversion cavity is inevitably longer than the optimum value.  This 
longer 
cavity apparently results in smaller effective magnetic field
$B_{\rm eff}$ due
to the weak magnetic field near the detection cavity region.
However actually this does not mean any serious deterioration of the 
detection sensitivity, since this effect of longer cavity has been 
taken into account from the first design stage. 

The only drawback of this longer cavity is that we have many series of 
higher TE modes which cross to the fundamental TM mode as we tune the 
frequency of the cavity, thus loosing many frequency bands due to the 
avoided crossings.  By using another rod with different diameter, 
however, we can fill these empty space in the frequency sweep, since the 
different rod brings avoid crossings at different frequencies.

\subsection{Cooling \protect\\ }
\label{sec:performance2}

\begin{figure}
\epsfig{file=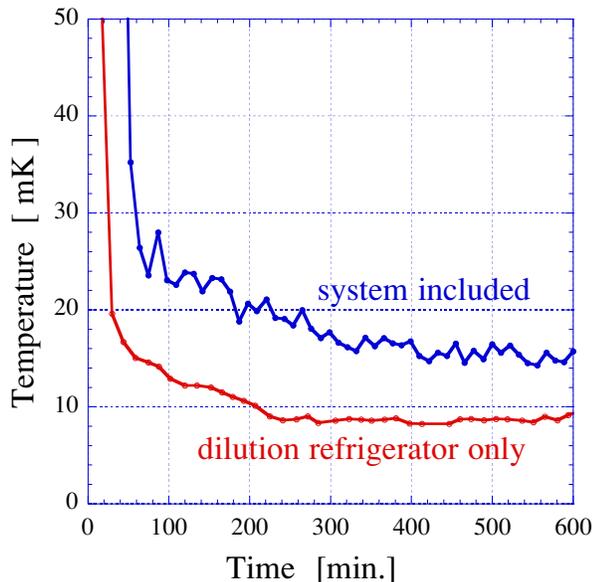, width=8cm}
\caption{Cooling rate of the present cavity system with a dilution 
refrigerator.  The time zero represents when the roots pump is on.
}
\label{fig:cooling}
\end{figure}

The temperature at the mixing chamber and the cavity was measured with
the anisotropy distribution of gamma rays from an oriented $^{60}$Co
single crystal source.  A thermometer of
RuO$_{2}$ was also used for the temperature measurements at several
places on the cavity surface.
The weight of the cavity system is about 25 kg and it takes
sometime to cool the whole cavity system down enough to 10 mK range.
In Fig.~\ref{fig:cooling} shown is the cooling characteristics of the cavity
system with time after the pumping of the $^3$He/$^4$He mixture gas was
started.  It takes about 5 hours for the whole system to be cooled
down to 12 mK.
%
%
%   eps figure   *******************************
%   eps figure   *******************************
%

\section{Discussion\protect\\ }

The system has fairly satisfactory performance for the present
purpose of the axion search experiment.  The temperature achieved for
the cavity is 12 $\sim$ 15 mK.  This could be further improved by reducing the
heat leak from the higher temperature side. 

The estimated Q value of the cavity from the surface resistance of the 
materials used is approximately given by 

\begin{equation}
Q_{c} \sim 1.3 \times 10^{5} \left( \frac{f_c}{1 {\rm GHz}} \right)^{-2/3}.   
\label{eqn:unloadQ}
\end{equation}

The actual Q achieved in the present system is somewhat lower than the value
estimated from this tendency. This lower Q value may be
due to many holes drilled and the associated equipment installed  in 
the cavity such 
as the Stark plate electrodes in the cavity.  Detailed analyses on this 
point, however, have not yet been tried.

The form factor $G$ of the present
magnet and the cavity system seems to be rather small even at the lower 
frequency region where the insertion length of the rod is rather small. 
As discussed already,  this lower value is, however, 
mainly due to the long length of the cavity, in which the magnetic
field at the lower section is quite small, because of the effect of
the cancellation magnet. If we neglect the contribution of the lower
part of the cavity in the evaluation of the effective magnetic field,
then the resulting values of $G$ are improved as shown also in 
Fig.~\ref{fig:fvsrod}.  

Further improvement on the value of the form factor may be
possible by adopting the metal/dielectric post system in replace of the
present rods.  Unfortunately this improvement is difficult in the
present structure of the cryostat, because no useful space is
available at the mixing chamber region where we have to install a
selective field ionization detector and microwave power input
circuitry to the cavity.  However in a new large scale apparatus called
CARRACK2~\cite{Tada}, we developed a new cryostat system in
which the detection cavity is set at the lower side and the laser and
the atomic beam are introduced horizontally through the lower part of
the vacuum chamber, thus enabling us to install the post driving
system for the tuning of the cavity frequency.
This development will be reported elsewhere.

\section{Summary\protect\\ }

We have developed the coupled cavity system for the dark matter axion search
with the Rydberg-atom cavity detector.  In order to fulfill the
necessary ingredients of the Rydberg-atom cavity detector, one of the
component of the OFHC-copper cavity (conversion cavity) is permeated by the
strong magnetic field, while the other component (detection cavity) was
arranged to be free from the external
magnetic field.  The detection cavity is made of superconducting Nb,
with which the external magnetic field is expelled out to fulfill the
above requirement.

The conversion cavity is attached to the bottom plate of the mixing chamber
of the
dilution refrigerator and thus the whole system is cooled down to 10 mK
range. Great care was taken to reduce the heat load from the higher
temperature side through the tuning rods driven from the cryostat outside
at room temperature. Especially Kevlar string of low heat conductivity was
used to access the tuning rod from the outside.
The minimum temperature achieved is 12 mK, and the loaded Q
value obtained is $3.5 \sim 4.5 \times 10^{4}$ for the whole range of the
frequency tuning of about 15 \%.

\acknowledgments
The authors would like to thank A. Masaike for his continuous
encouragement throughout this work. They are also indebted to S.
Takeuchi of JAERI, Tokai for giving us invaluable information on the
Nb cavity fabrications.  This research was partly supported by a 
Grant-in-Aid for Specially Promoted Research (No. 09102010) by the 
Ministry of Education, Science, Sports, and Culture (Monbusho), Japan.

\end{document}